\documentclass[useAMS, usenatbib]{mn2e}
\usepackage{epsfig}
\usepackage{subfigure}
\usepackage{graphicx}
\usepackage{rotating}
\author[Spreckley S. A., Stevens. I. R.]{Spreckley S. A.$^{1}\thanks{E-mail:
sas@star.sr.bham.ac.uk, irs@star.sr.bham.ac.uk}$, Stevens. I. R.$^{1}$\\
$^{1}$School of Physics and Astronomy, University of Birmingham, Edgbaston, Birmingham B15 2TT}

\title[Period and amplitude of Polaris]{The period and amplitude changes of Polaris ($\alpha$ UMi) 
from 2003 to 2007 measured with SMEI}
\date{Accepted 2008 May 8}

\begin{document}

\maketitle

\begin{abstract}
We present an analysis of 4.5 years of high precision (0.1\%) space-based
photometric measurements of the Cepheid variable Polaris, obtained by
the broad band Solar Mass Ejection Imager (SMEI) instrument on board the Coriolis
satellite.  The data span from April 2003 to October 2007, with a 
cadence of 101 minutes and a fill factor of 70\%.  We have measured the mean peak to peak 
 amplitude across the whole set of observations 
 to be 25 mmag.  There is, however, a clear trend that the size of the oscillations has been increasing
 during the observations, with peak to peak variations less than 22 mmag in 
 early 2003, increasing to around 28 mmag by October 2007, suggesting that the peak to peak 
 amplitude is increasing at a rate of  $1.39 \pm 0.12$~mmag yr$^{-1}$.  Additionally, 
 we have combined our new measurements with archival measurements to measure a
 rate of period change of $4.90\pm 0.26$~s yr$^{-1}$ over the last 50~years.  However,
 there is some suggestion that the period of Polaris has undergone a recent
 decline, and combined with the increased amplitude, this could imply evolution
 away from an overtone pulsation mode into the fundamental or a double pulsation mode depending
 on the precise mass of Polaris.
\end{abstract}


\section{Introduction}
In spite of Julius Caesar's view (Shakespeare 1623, Act 3 Scene I),
Polaris is in fact one of the more inconstant of stars. In addition to
not being precisely located at the North Celestial Pole, it is a
variable star, with a pulsation period of nearly 4 days and a current
pulsation amplitude of around 30-50 mmag in the V band. Additionally, 
Polaris is not even constant in its inconstancy, as both the pulsation 
period and the pulsation amplitude have changed in the past.
 The amplitude in particular has changed substantially and is currently still
 changing, as we will describe in this paper.

Polaris is an important star for a number of reasons -- it is the
nearest Cepheid variable and a star where we can see stellar evolution
taking place. Understanding the location of Polaris on the
Hertzsprung-Russell diagram (and particularly the relationship of
Polaris to the Instability Strip; whether it is a star undergoing its
first or third or even fifth crossing) and the nature of the
pulsations (whether fundamental model or an overtone pulsator) are
all important questions in stellar evolution.

The first evidence suggesting the variable nature of Polaris was presented
150 years ago \citep{Seidel1852,Schmidt1857} with strong confirmation, along with 
the correct period, being supplied by \citet{Camp99} via radial velocity measurements. 
 Photometric detection of the pulsations were presented several years afterwards
 \citep{Hertz11,Pannekoek13}.  A large number
 of observations have have been made in the intervening years, which have
 helped to build a picture of how the star has been evolving over the last
 century and a half.  Despite this, there is still a great deal of interest
 in Polaris due to the changes in the period and amplitude of the oscillations,
 as well as unusual events such as the change from a steady decline in amplitude to 
 a very rapid decline from 0.1 mag to $\sim$0.02 mag during the 1960's. During this
 time, the period also readjusted downwards.

The rate of period change is an important diagnostic tool for determining which 
crossing of the instability strip a Cepheid 
 is undergoing.  The recent analysis of the O-C residuals of Polaris by \citet{Turner05} has
 led to the suggestion that the period of Polaris is currently increasing
 at a rate of 4.5 s yr$^{-1}$.  This rate of period change is unusual for 
 a Cepheid with this period and adds to the confusion as to what stage of
 its evolution Polaris is actually at.
 In the following sections we present new SMEI photometric observations of Polaris,
 discuss the amplitude changes that we observe during the course of the
 observations and also look at the O-C residuals and interpret these in
 the light of recent measurements of the period.

\begin{figure*}
\begin{center}
\begin{sideways}
\epsfxsize=7cm\epsfbox{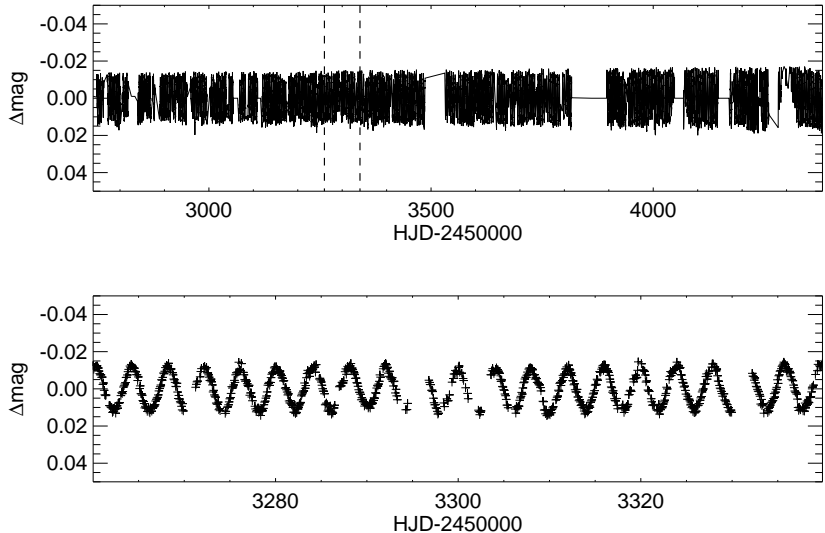}
\end{sideways}
\caption{The complete 4.5 year SMEI time-series of Polaris is shown in the upper 
 panel, whilst a short section from 13th September to 29th  November can be seen 
 in the lower panel.  The relevant section of the full time series that has been 
 depicted in the lower panel, is highlighted by the vertical dashed lines.\label{timeseries}}
\end{center}
\end{figure*}

\section{Data preparation and Analysis}

The new photometry we present was obtained using the SMEI instrument
on board the USAF Coriolis spacecraft, which was launched in January
2003 into an 840~km Sun-synchronous polar orbit with
an orbital period of 101 minutes. SMEI consists of  3 cameras, each with a
field of view of $60^{\circ} \times 3^{\circ}$, which monitors nearly the
entire sky over one orbit. Consequently, we obtain data for Polaris on
essentially every orbital pass.  SMEI has a roughly triangular pass
 band with a peak quantum efficiency of 47\% at 700 nm and falling to
 5\% at 430 nm and 1025 nm.  Although SMEI is a small instrument,
the fact that it has monitored the entire sky with a cadence of $\sim
100$ minutes for over 4 years, results in stellar light curves, for
bright stars, that are unprecedented.  An overview of the SMEI instrument 
can be found in \citet{Eyles03}, and an overview of stellar variability 
results being obtained with SMEI will be presented in \citet{Spreck08}.
SMEI results on the variability of the Red Giant Arcturus 
can be found in \citet{Tarrant07}.

The Polaris data spans from April 2003 to October 2007, with a 70\%
fill throughout this period of time, giving us an exceptional data
sample to investigate the period and amplitude variations of the
3.97 day oscillations exhibited by Polaris.  The full details of the 
reduction pipeline for generating time-series
from the SMEI data will be presented in \citet{Spreck08}, so we only
discuss the data reduction briefly here.  

The raw images obtained by the SMEI instrument are bias subtracted, 
 have a temperature scaled
dark current signal removed, and are flat fielded.  Hot pixels and
high energy particle hits are corrected on the images via
interpolation, before aperture photometry is performed.  The resulting
light curves are corrected for systematic effects resulting from the
variation of the PSF as it moves across the CCD and vignetting/optical
effects.  Removing a best fit sine curve from the entirety of the dataset
 highlighted non-regular systematic variations at the few mmag level which
 we have largely removed using a smoothed box car average 
 obtained with a window width of $\sim$12 days.  We finally removed a number of spurious data 
 points from the data, which were primarily due to uncorrected cosmic rays, 
by performing a 3 sigma clip on short 28 day sections of data from which the best
 fitting sinusoidal relation for each section had been removed.

The resulting time series can be seen in Fig. \ref{timeseries} along with a 
closer view of a section of data taken from 13th September 2006 to 29th November 2006,
 which highlights the level of precision we are able to attain over a long baseline.

\section{Results}
\begin{figure}
\begin{center}
\begin{sideways}
\epsfxsize=5.8cm\epsfbox{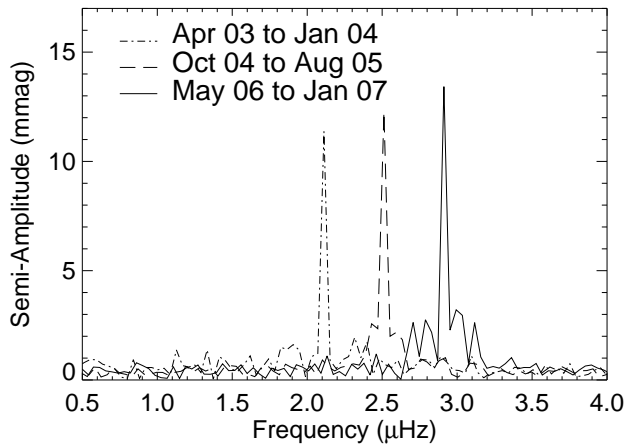}
\end{sideways}
\caption{The Fourier spectra for 3 separate sections of the data, each spanning
 approximately 9 months.  The spectra for April 2003 to January 2004, and October 
 2004 to August 2005 have been offset by 0.8 and 0.4 $\mu$Hz respectively from the
 2.914 $\mu$Hz peak from the May 2006 to January 2007 data, to show the increasing 
amplitude.  We find no evidence of a weak fundamental mode at $\sim 2.1\mu$Hz
 above the noise level in these spectra, with an amplitude above 1mmag.}\label{three_epoch}
\end{center}
\end{figure}

\begin{figure}
\begin{center}
\begin{sideways}
\epsfxsize=5.75cm\epsfbox{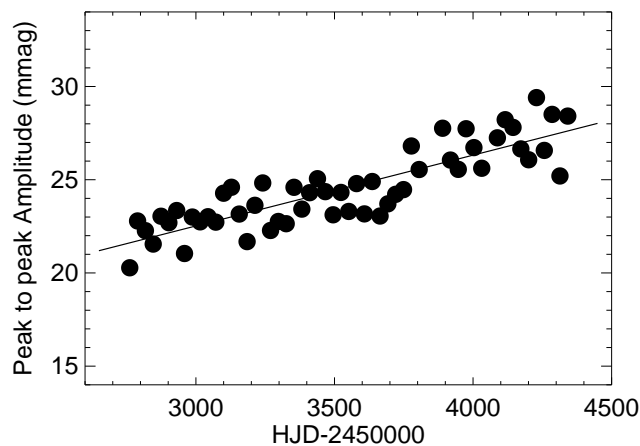}
\end{sideways}
\caption{The mean peak to peak amplitude of the 2.914$\mu$Hz (3.97 day) oscillations has risen at a rate of 
$1.39\pm 0.12$ mmag~yr$^{-1}$ over the last 4.5 years.  The scatter exhibited about the mean 
trend is in part due to the fact that the amplitude of the oscillations of Polaris vary from 
cycle to cycle.\label{monthly_FFT}}
\end{center}
\end{figure}

In order to study the amplitude of the 3.97 day (2.914$\mu$Hz) oscillation we have generated
Fourier spectra of both long (8-9 months) and short (28 day) sections
of the time series data. The resulting trend from computing the mean
 amplitude in each data chunk is shown in Fig.~\ref{three_epoch}
 and Fig.~\ref{monthly_FFT}.  To create the Fourier spectra, we first subtracted a  
 robust mean from the time series and the residuals were then converted
 to a change in magnitude relative to the mean magnitude.  
 After computing the Fourier spectra we ensured that in every case
 the power corresponding to the 3.97 day oscillation was restricted to
 a single bin, and padding the data with zeros to artificially enhance
 the resolution of the Fourier spectra had no effect on the computed
 amplitudes.

Fig.~\ref{three_epoch} shows the 2.914 $\mu$Hz peak in the Fourier spectra for 
three $\sim$9 month sections of the light curve.  The peaks for the 2003 and 2004-2005 data have been
 offset by $-$0.8 and $-$0.4 $\mu$Hz respectively to highlight the increase
 in the amplitude over time.  On the assumption that Polaris is oscillating in the first 
 overtone mode (see section 4), we do not detect any significant power at the expected frequency
 for the fundamental mode, i.e. $\sim 2.1\mu$Hz.
 The trend of the increasing amplitude is highlighted more clearly in 
 Fig.~\ref{monthly_FFT} where we have plotted the mean amplitude calculated in consecutive 28 day
 chunks of data. Each 28 day section contained 400 data points after data gaps had been filled with 
 zero values.  The best fitting linear relation 
 describing the rate of increase for the peak to peak amplitude is $1.39\pm 0.12$ mmag yr$^{-1}$.  
 Some of the scatter in the plot is attributed to the varying amplitude of the oscillations,
 which \citet{Evans04} also saw in WIRE observations, and suggested it could be due to the analogue 
of the Blazkho effect in Cepheids.  In our analysis, however, we do not see any significant periodic variations in
 the amplitude above a level of $\sim1$~mmag appearing consistently throughout the observations. 

This is to our knowledge the first highly confident detection of the amplitude increase over the
 last few years from photometric measurements, although some hint has been given previously
 \citep{Davis03, Engle04}.  This completely contradicts the claim that
  Polaris is about to cease its variability and leave the instability strip \citep{Dinshaw89}, 
 although this was based somewhat on an erroneous result.

\begin{table}
\caption{Recent measurements of the period of Polaris suggest that it has undergone another recent decline. 
 There seems to have been a large decline between 1988 and 1993, with the rate slowing in the last ten years.
 In total, the period has reduced by around 200 seconds over the last 20 years.\label{periods}}
\begin{center}
\begin{tabular}{|c|l|l|}
\hline
Year & Period  & Reference\\
     & (days) & \\
\hline
1987-1988 & $3.9746\pm 0.0008$  & \citet{Dinshaw89} \\
1993-1994 & $3.97268\pm 0.00011$ & HC00            \\
1994-1997 & $3.972352\pm 0.000003$ & HC00          \\
          &                         & \citet{KF98} \\
2003-2007 & $3.97209\pm 0.00004$ & This paper       \\
\hline
\end{tabular}
\end{center}
\end{table}

\begin{figure}
\begin{center}
\epsfxsize=7cm\epsfbox{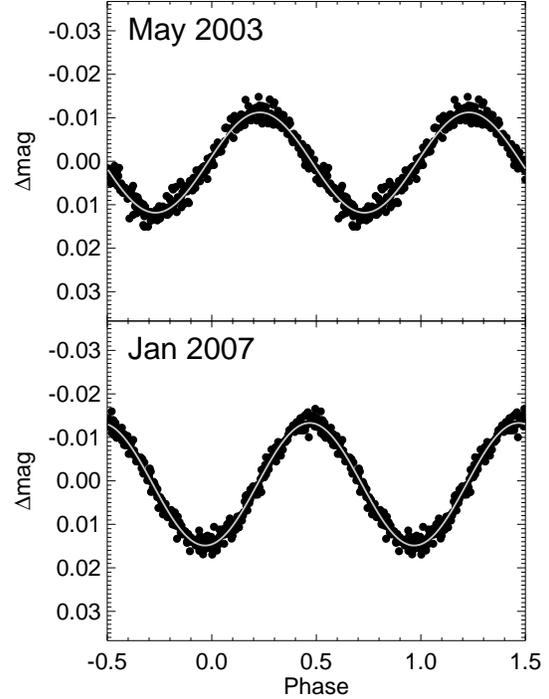}
\vspace{0.2cm}
\caption{The phase folded light curves containing data from May 2003, and January 2007 highlight
 the excellent quality of data we have for computing the O-C residuals.  One can clearly discern 
 the increase in amplitude as well as the phase offset due to the changing period of Polaris,
 between the two light curves.\label{phased_lc}}
\end{center}
\end{figure}

O-C residuals have been computed for each of the 28 day sections also.
New times and phases of light maximum were determined by using the mean 
amplitudes calculated in the previous step to perform least squares fitting of
 the data, which has been re-phased to the period and epoch presented
 in \citet{Berd95}:
\begin{equation}
 \textrm{HJD}_{\textrm{max}} = 2,428,260.727 + 3.969251E
\end{equation}
\noindent where $E$ is the number of elapsed cycles since this epoch.  Examples
 of the phase folded data used to determine the O-C residuals are shown in
 Fig.~\ref{phased_lc}.  In this figure, the increase in amplitude over time, 
and the changing phase offset can be clearly discerned.  The full set of
 O-C residuals are listed in Table~\ref{OCtab}.

\begin{table}
\caption{Measurements of the times of maxima for the oscillations of Polaris, determined from
 fitting 1 month sections of data.  N is the number of points used to compute each 
O-C value.\label{OCtab}}
\begin{center}
\begin{tabular}{|c|c|c|c|}
\hline
     HJD     &  Cycle  & O-C & N \\
\hline
 2452763.768 & 6171 & $8.793 \pm  0.055$ & $198$ \\
 2452787.637 & 6177 & $8.846 \pm  0.040$ & $311$ \\
 2452815.417 & 6184 & $8.842 \pm  0.071$ & $170$ \\
 2452855.113 & 6194 & $8.845 \pm  0.063$ & $235$ \\
 2452870.954 & 6198 & $8.810 \pm  0.085$ & $182$ \\
 2452902.800 & 6206 & $8.901 \pm  0.036$ & $333$ \\
 2452930.578 & 6213 & $8.894 \pm  0.040$ & $310$ \\
 2452962.393 & 6221 & $8.956 \pm  0.051$ & $220$ \\
 2452986.213 & 6227 & $8.960 \pm  0.046$ & $271$ \\
 2453017.990 & 6235 & $8.983 \pm  0.039$ & $302$ \\
 2453041.822 & 6241 & $8.999 \pm  0.032$ & $252$ \\
 2453077.577 & 6250 & $9.031 \pm  0.035$ & $205$ \\
 2453097.476 & 6255 & $9.084 \pm  0.030$ & $283$ \\
 2453133.211 & 6264 & $9.096 \pm  0.037$ & $295$ \\
 2453157.013 & 6270 & $9.083 \pm  0.048$ & $344$ \\
 2453184.832 & 6277 & $9.116 \pm  0.053$ & $305$ \\
 2453212.657 & 6284 & $9.157 \pm  0.046$ & $287$ \\
 2453240.456 & 6291 & $9.171 \pm  0.064$ & $307$ \\
 2453268.224 & 6298 & $9.154 \pm  0.031$ & $333$ \\
 2453296.050 & 6305 & $9.196 \pm  0.044$ & $275$ \\
 2453323.862 & 6312 & $9.223 \pm  0.036$ & $326$ \\
 2453355.648 & 6320 & $9.254 \pm  0.047$ & $299$ \\
 2453383.429 & 6327 & $9.251 \pm  0.038$ & $335$ \\
 2453411.268 & 6334 & $9.305 \pm  0.035$ & $280$ \\
 2453439.071 & 6341 & $9.324 \pm  0.034$ & $353$ \\
 2453466.878 & 6348 & $9.346 \pm  0.034$ & $319$ \\
 2453482.754 & 6352 & $9.345 \pm  0.087$ & $ 73$ \\
 2453534.432 & 6365 & $9.422 \pm  0.067$ & $ 61$ \\
 2453550.291 & 6369 & $9.404 \pm  0.032$ & $276$ \\
 2453582.042 & 6377 & $9.401 \pm  0.032$ & $354$ \\
 2453609.845 & 6384 & $9.420 \pm  0.039$ & $282$ \\
 2453633.720 & 6390 & $9.480 \pm  0.036$ & $290$ \\
 2453661.498 & 6397 & $9.473 \pm  0.029$ & $306$ \\
 2453693.278 & 6405 & $9.498 \pm  0.029$ & $352$ \\
 2453721.077 & 6412 & $9.512 \pm  0.029$ & $353$ \\
 2453748.891 & 6419 & $9.542 \pm  0.027$ & $285$ \\
 2453776.690 & 6426 & $9.556 \pm  0.027$ & $307$ \\
 2453808.445 & 6434 & $9.557 \pm  0.033$ & $170$ \\
 2453899.794 & 6457 & $9.614 \pm  0.127$ & $114$ \\
 2453915.709 & 6461 & $9.651 \pm  0.044$ & $347$ \\
 2453947.502 & 6469 & $9.690 \pm  0.034$ & $291$ \\
 2453975.325 & 6476 & $9.729 \pm  0.043$ & $343$ \\
 2453999.122 & 6482 & $9.710 \pm  0.035$ & $225$ \\
 2454030.908 & 6490 & $9.742 \pm  0.028$ & $373$ \\
 2454086.529 & 6504 & $9.793 \pm  0.030$ & $349$ \\
 2454118.265 & 6512 & $9.775 \pm  0.028$ & $323$ \\
 2454138.145 & 6517 & $9.809 \pm  0.033$ & $208$ \\
 2454177.925 & 6527 & $9.896 \pm  0.040$ & $147$ \\
 2454197.761 & 6532 & $9.886 \pm  0.031$ & $343$ \\
 2454225.598 & 6539 & $9.939 \pm  0.104$ & $268$ \\
 2454253.344 & 6546 & $9.900 \pm  0.083$ & $207$ \\
 2454289.133 & 6555 & $9.966 \pm  0.071$ & $125$ \\
 2454316.910 & 6562 & $9.958 \pm  0.053$ & $223$ \\
 2454340.754 & 6568 & $9.987 \pm  0.026$ & $342$ \\

\hline
\end{tabular}
\end{center}
\end{table}

\begin{figure}
\begin{center}
\begin{sideways}
\epsfxsize=5.8cm\epsfbox{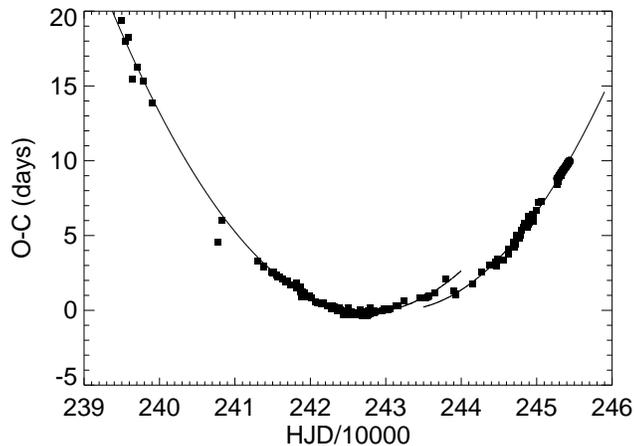}
\end{sideways}
\vspace{0.3cm}
\caption{The O-C diagram for Polaris. We have combined SMEI results with archival data (see \citet{Turner05} 
 for a full list of references).
The best fit parabola for the pre-1963 data is as in \citet{Turner05}, the parabolic fitting to the post 
1965 data, including our new results at the right of the plot, gives a period change of
 $4.90\pm 0.26$~s~yr$^{-1}$.\label{OC}}
\end{center}
\end{figure}

The O-C residuals obtained from \citet{Turner05}, along with the new values calculated from our data
are plotted in Fig.~\ref{OC}.  We have used the same time regimes as this paper (i.e. pre-1963
 and post 1965) to determine our rate of period change.  The best fitting parabolic relation for 
 observations before 1963, as determined by \citet{Turner05}, provides an estimate for the rate of period 
 increase over this time of $4.44\pm 0.03$ s yr$^{-1}$   The best fitting parabolic relation for the data since 1963,
 with the inclusion of our new measurements suggests the mean rate of increase for the period has increased 
 to $4.90\pm 0.26$ s yr$^{-1}$   This relation is again shown in Fig.~\ref{OC}.  Ignoring the data from 
 1966, as in \citet{Turner05}, insignificantly alters the value to $4.99\pm 0.29$ s yr$^{-1}$.  Ignoring the 
 datum from 1965, however, causes a dramatic change to the calculated value, giving instead 
 $4.46\pm 0.32$ s yr$^{1}$.  It is clear therefore that the mean rate of period increase over the last
  50 years has been between 4.4 and 5 s yr$^{-1}$, consistent with the rate before 1963.  

If one looks more closely at recent measurements for the period, however, it does appear that it may have recently 
 undergone a rapid decline, similar to that seen in the early 1960's.  Table~\ref{periods} shows the period as 
 measured several times over the last 20 years with the values obtained from \citet{Hatzes00} and references 
therein, together with 
 the period measured from our new results.  The decline is very evident, and amounts to a decrease of around 200 
 seconds during the last 20 years, but the rate has been much slower over the last ten years than it was between
 1987 and 1997.  Additionally, although our results are fairly consistent with a period increase
 of 4.9~s~yr$^{-1}$, they do follow a slightly shallower trend which is likely due to the recent decrease
 in period.
 
We will now discuss these results in the context of stellar evolution and ascertain what implications they have on 
the evolutionary stage of Polaris.

\section{Discussion}
There is a great deal of evidence to suggest that Polaris is a first overtone (s-Cepheid) oscillator.
  \citet{Feast97} used the Hipparcos parallax (measured to be $7.56\pm 0.48$~mas for Polaris) to fit
 Period-Luminosity models to a sample of Cepheids and concluded that the best fit for Polaris
 resulted if it was treated as a first overtone pulsator.  The updated value for the Hipparcos
 parallax of Polaris is $7.54\pm 0.09$~mas \citep{vanL07a,vanL07b}, therefore this conclusion
 is still valid.  \citet{Nord99} used interferometry to measure the radius of Polaris to be $46\pm 3 R_{\sun}$, 
 and again this is only consistent with the pulsation period if Polaris is a first overtone pulsator.
 The mean rate of change of the period and the small amplitude of the oscillations are also
 indicators that Polaris oscillates in an overtone mode.  Combining this with the fact that Polaris
 exhibits a highly symmetrical light curve, it exhibits all of the features we expect from s-Cepheids.

\begin{figure}
\begin{center}
\begin{sideways}
\epsfxsize=5.65cm\epsfbox{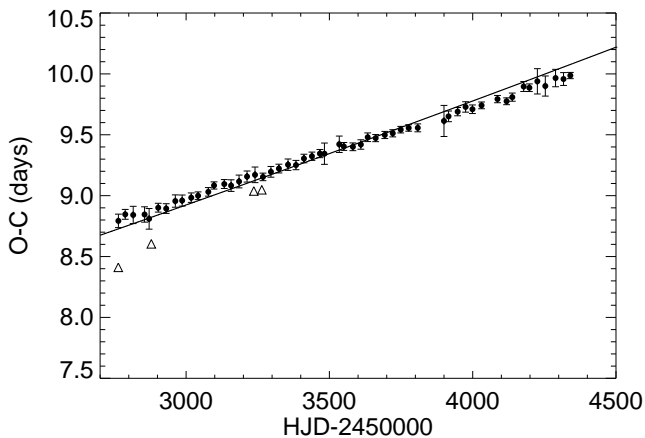}
\end{sideways}
\caption{The O-C data obtained using SMEI, depicted by filled circles with error
 bars, along with the measurements from \citet{Turner05}, shown by triangles. Our
 measurements are fairly consistent with the 4.90~s~yr$^{-1}$ period change, but do
 appear to follow a slightly shallower trend.
\label{SMEI_OC}}
\end{center}
\end{figure}

\citet{Evans02} suggest that Polaris exists at the cool edge of the region of the 
 instability strip occupied by the s-Cepheids, but that the positive period change 
 could not be due to evolution as the star would be evolving towards the centre of the instability strip, 
 which would defy the previously declining amplitude. \citet{Turner05} also place Polaris on the 
 red edge of the instability strip for putative first crossers, which corresponds to the s-Cepheid red edge 
 in this case.  They do however suggest the possibility of Polaris being a fundamental pulsator, which is 
 unlikely given the evidence above.  \citet{Dinshaw89} on the 
 other hand believed that Polaris was about to evolve out of the instability strip completely, 
 which would require it to be near the edge of the instability strip, which it certainly does not appear to be.  
 The behaviour we are now seeing from Polaris is consistent with what we expect from a Cepheid located in the
 instability strip where \citet{Evans02} and \citet{Turner05} suggest, but the conclusion that the period change is 
 not due to evolution may be incorrect.

\begin{figure}
\begin{center}
\epsfxsize=7.9cm\epsfbox{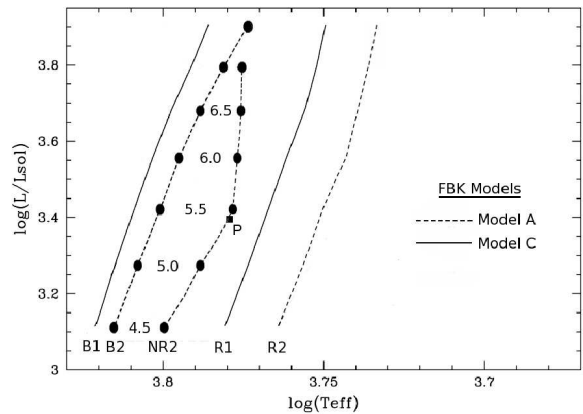}
\caption{Theoretical models for the first overtone Cepheid instability strip (IS), computed by FBK.
 Results for convective model A (dashed lines) and convective model C (solid lines) are plotted.  
 The labels B1 and B2 refer to the linear model blue edges for models A and C respectively, whilst R1 and R2 refer 
 to the red edges.  NR2 refers to the non-linear model red edge, and the values represent the masses used
 in the models.  Plotting Polaris on the IS (filled square marked by P), we see it lies on the 
 non-linear red edge boundary.\label{ISmodel}}
\end{center}
\end{figure}

 Firstly, consider the evidence that the period has once again undergone a rapid decline, which could be a phase of
 blueward evolution, as \citet{Turner05} suggests could be the case for the 1963-66 period
 readjustment.  Secondly, the amplitude appeared to cease its decline in the early 1990s, and
 is now seemingly increasing again.  One might expect to see such behaviour if Polaris was undergoing
 an evolutionary change in its oscillation mode.  If we look at the models for s-Cepheids produced by
 \citet{FBK00},  and specifically look at the location of Polaris in the computed 
 instability regimes then,
 as Fig.~\ref{ISmodel} shows, Polaris lies on the red edge of the overtone instability strip for their non-linear
 convective model, which is able to generate light and radial velocity curves very similar to those
 found observationally.  If the true red edge of the first overtone instability strip lies close to the one computed 
 in this FBK model, then Polaris is in fact undergoing a change from being a first overtone pulsator (assuming
 it is evolving to the cooler side of the instability strip) to becoming
 either a fundamental pulsator or a double mode pulsator.  The position of Polaris in Fig.~\ref{ISmodel} suggests
 that its mass is not quite great enough to enter the fundamental pulsator regime, which occurs
 for masses greater than around $5.5~M_{\sun}$ (the knee in the model track), but recent measurements by 
 \citet{Evans07} for example, do not place enough of a constraint on the mass to make an absolute determination as to 
 which regime Polaris will enter.  If Polaris is about to cross into another pulsation regime, then we might
 expect to see occasional blips in the period as this readjustment phase takes place.

 Finally, the crossing mode of Polaris is also uncertain.  \citet{Turner05} suggest that despite Polaris 
 exhibiting a deficiency of carbon and an over-abundance of nitrogen \citep{BL81,LB86}, this cannot be
 interpreted in any fashion to determine the crossing mode, such as was done by \citet{KAUK96}.  Somewhat
 controversially, \citet{AKU94} suggests that Polaris has a small over-abundance
 of carbon, which would certainly place it in a first crossing scenario.

The rate of period change for Cepheids is an indicator of the crossing mode \citep{Turner06}.
  If we assume Polaris to be oscillating in the first overtone mode, and take the rate 
of period change over the last 150 years to be +4.5 s yr$^{-1}$, then this is a factor of $\sim$3 
too small for a first crossing Cepheid and about the same factor too large for a third crossing 
Cepheid.  \citet{Turner05} suggest that the observable characteristics of Polaris are most consistent
 with a first time crosser.  Interestingly, however, if one only considers the trend in the period from 
the last 20 years, 
then we see a decline at a rate of $\sim10$ s yr$^{-1}$.  This is actually consistent with a first 
overtone Cepheid undergoing its second crossing.  The possibility that Polaris is in its second 
crossing was also discussed by \citet{Engle04}.  We must be cautious, however, as the rate over the last
 fourteen years, using measurements with much better precision than those presented in \citet{Dinshaw89},
 only suggest a decline of $\sim3$ s yr$^{-1}$.  Clearly, it is difficult to draw any firm conclusions 
on the crossing mode at present.

\section{Conclusions}
The new results obtained with SMEI strongly suggest the amplitude of Polaris is once again increasing.
  The star does seem to be oscillating in the first overtone mode, but likely lies close to the red
 edge of the instability strip for s-Cepheids, and is therefore likely to soon evolve into either a fundamental
 or double-mode pulsation Cepheid, assuming the Cepheid is undergoing its first or third crossing.  It is 
uncertain how quickly Polaris will evolve across the boundary between pulsation regimes and what behaviour
 the Cepheid will exhibit as it does so.  One slight oddity in the results is the lack of evidence for the
 fundamental mode of oscillation being present, but this may appear in the near future.  It is therefore 
 crucial that high precision monitoring of this star is continued for the foreseeable future so that the 
changes can be watched closely.

\section{acknowledgements}
We would like to thank the referee, Clifton Laney, for his extremely useful comments, which have helped
 improve this paper.  SAS acknowledges support from STFC and the School of Physics and Astronomy, 
University of Birmingham.   SMEI was designed and built by members of UCSD, AFRL, and the University 
of Birmingham. We particularly  thank Yvonne Elsworth, Andrew Buffington, Chris Eyles and James Tappin.


\begin{thebibliography}{}

\bibitem[\protect\citeauthoryear{Andrievsky, Kovtyukh \& Usenko}{Andrievsky et~al.}{1994}]{AKU94}
Andrievsky S.M., Kovtyukh V.V., Usenko I.A., 1994, A\&A, 281, 465

\bibitem[\protect\citeauthoryear{Berdnikov \& Pastukhova}{Berdnikov \& Pastukhova}{1995}]{Berd95}
Berdnikov L.N., Pastukhova E.N., 1995, Astron. Lett., 21, 369

\bibitem[\protect\citeauthoryear{Boyarchuk \& Lyubimkov}{Boyarchuk \& Lyubimkov}{1981}]{BL81}
Boyarchuk A.A., Lyubimkov L.S., 1981, Comm. Crimean Astrophys Obs., 63, 66.

\bibitem[\protect\citeauthoryear{Campbell}{Campbell}{1899}]{Camp99}
Campbell W. W., 1899, ApJ, 10, 180

\bibitem[\protect\citeauthoryear{Cox}{Cox}{1998}]{Cox98}
Cox A.N., 1998, ApJ, 496, 246

\bibitem[\protect\citeauthoryear{Dinshaw~et~al.}{Dinshaw~et~al.}{1989}]{Dinshaw89}
Dinshaw A.N., Matthews J.M., Walker G.A.H., Hill, G.M., 1989, AJ, 98, 2249

\bibitem[\protect\citeauthoryear{Davis~et~al.}{Davis~et~al.}{2003}]{Davis03}
Davis J.J., Tracey J.C., Engle S.G., Guinan E.F., 2003, BAAS, 34, 1296

\bibitem[\protect\citeauthoryear{Engle, Guinan \& Koch}{Engle et al.}{2004}]{Engle04}
Engle S.G., Guinan E.F., Koch R. H., 2004, BAAS, 36, 744

\bibitem[\protect\citeauthoryear{Evans, Sasselov \& Short}{Evans et~al.}{2002}]{Evans02}
Evans N.R., Sasselov D.D., Short C.I., 2002, ApJ, 567, 1121

\bibitem[\protect\citeauthoryear{Evans et~al.}{Evans et~al.}{2004}]{Evans04}
Evans N.R., Buzasi D.,  Sasselov D.D., Preston H., 2004, BAAS, 36, 1429

\bibitem[\protect\citeauthoryear{Evans et~al.}{Evans et~al.}{2007}]{Evans07}
Evans N.R. et~al., 2007, IAUS, 240, 102

\bibitem[\protect\citeauthoryear{Eyles et al.}{Eyles et~al.}{2003}]{Eyles03} 
Eyles C.~J., et al.\ 2003, Solar Physics, 217, 319 

\bibitem[\protect\citeauthoryear{Feast \& Catchpole}{Feast \& Catchpole}{1997}]{Feast97} 
Feast M.W., Catchpole R.M., 1997, MNRAS, 286, L1

\bibitem[\protect\citeauthoryear{Feuchtinger, Buchler, \& Koll\'{a}th}{Feuchtinger et~al.}{2000, hereafter FBK}]{FBK00}
Feuchtinger M., Buchler J.M., Kollath Z., 2000, ApJ, 544, 1056 (FBK)

\bibitem[\protect\citeauthoryear{Hatzes \& Cochran}{Hatzes \& Cochran}{2000, hereafter HC00}]{Hatzes00}
Hatzes A.P., Cochran W.D., 2000, AJ, 120, 979 (HC00)

\bibitem[\protect\citeauthoryear{Hertzsprung}{Hertzsprung}{1911}]{Hertz11}
Hertzsprung E., 1911, Astron. Nach., 189, 89

\bibitem[\protect\citeauthoryear{Kamper \& Fernie}{Kamper \& Fernie}{1998}]{KF98}
Kamper K.W., Fernie J.D., 1998, AJ, 116, 936

\bibitem[\protect\citeauthoryear{Kovtyukh et~al.}{Kovtyukh et~al.}{1996}]{KAUK96}
Kovtyukh V.V., Andrievsky S.M., Usenko I.A., Klochkova V.G., 1996, A\&A, 316, 155

\bibitem[\protect\citeauthoryear{Luck \& Bond}{Luck \& Bond}{1986}]{LB86}
Luck R.E., Bond H.E., 1986, PASP, 98, 442

\bibitem[\protect\citeauthoryear{Nordgren et~al.}{Nordgren et al.}{1999}]{Nord99}
Nordgren T.E. et~al., 1999, AJ, 118, 3032

\bibitem[\protect\citeauthoryear{Pannekoek}{Pannekoek}{1913}]{Pannekoek13}
Pannekoek A., 1913, Astron. Nachr., 194, 359

\bibitem[\protect\citeauthoryear{Seidel}{Seidel}{1852}]{Seidel1852}
Seidel L., 1852, IIK1., Dokl. Akad. Wiss. Munchen, 6, 564

\bibitem[\protect\citeauthoryear{Schmidt}{Schmidt}{1857}]{Schmidt1857}
Schmidt J.F.J., 1857, Astron. Nachr., 46, 293

\bibitem[\protect\citeauthoryear{Simon \& Lee}{Simon \& Lee}{1981}]{Simon81}
Simon N.R., Lee A.S., 1981, ApJ, 248, 291

\bibitem[\protect\citeauthoryear{Spreckley \& Stevens}{Spreckley \& Stevens}{2008}]{Spreck08}
Spreckley S.A., Stevens I.R., 2008, in prep

\bibitem[\protect\citeauthoryear{Tarrant et~al.}{Tarrant et~al.}{2007}]{Tarrant07} 
Tarrant N.~J., Chaplin W.~J., Elsworth Y., Spreckley S.~A., Stevens I.~R.\ 2007,  MNRAS, 382, L48 

\bibitem[\protect\citeauthoryear{Turner et~al.}{Turner et~al.}{2005}]{Turner05}
Turner D.G., Savoy J., Derrah J., Abdel-Sabour Abdel-Latif M.,  Berdnikov, L.N., 2005, PASP, 117, 207

\bibitem[\protect\citeauthoryear{Turner et~al.}{Turner et~al.}{2006}]{Turner06}
Turner D.G., Abdel-Sabour Abdel-Latif M., Berdnikov L.N., 2006, PASP, 118, 410

\bibitem[\protect\citeauthoryear{van Leeuwen}{van Leeuwen}{2007}]{vanL07a}
van Leeuwen F., 2007,  Hipparcos, the New Reduction of the Raw Data, Kluwer Academic Publishers

\bibitem[\protect\citeauthoryear{van Leeuwen et~al.}{van Leeuwen et~al.}{2007}]{vanL07b}	
van Leeuwen F.,  Feast M.W., Whitelock P,A., Laney C.D., 2007, MNRAS, 379, 723

\end{thebibliography}
\end{document}